\documentclass[conference]{IEEEtran}
\IEEEoverridecommandlockouts
\usepackage{cite}
\usepackage{amsmath,amssymb,amsfonts}
\usepackage{amsmath, amsthm}
\usepackage{fancyhdr}
\usepackage{algorithmic}
\usepackage{authblk}
\usepackage[ruled,linesnumbered,lined,boxed,commentsnumbered]{algorithm2e}
\usepackage{graphicx}
\usepackage{amsmath}
\usepackage{textcomp}
\usepackage{xcolor}
\usepackage{upgreek}

\newtheorem{condition}{Condition}
\newtheorem{remark}{Remark}

\newtheorem{theorem}{Theorem}
\newtheorem{lemma}{Lemma}
\usepackage{bm}
\usepackage[left=0.63in,top=0.7in,right=0.63in,bottom=0.95in]{geometry}

\usepackage{afterpage}

\def\BibTeX{{\rm B\kern-.05em{\sc i\kern-.025em b}\kern-.08em
    T\kern-.1667em\lower.7ex\hbox{E}\kern-.125emX}}

\begin{document}
\title{Resource Allocation for Semantic Communication under Physical-layer Security}
\author[1]{Yang~Li}
\author[1]{Xinyu~Zhou}
\author[2]{Jun Zhao}
\affil[1]{\small  School of Computer Science \& Engineering and ERI@N, Nanyang Technological University, Singapore}
\affil[2]{\normalsize School of Computer Science \& Engineering, Nanyang Technological University, Singapore} 
\affil[ ]{\normalsize 
yang048@e.ntu.edu.sg,
xinyu003@e.ntu.edu.sg,  junzhao@ntu.edu.sg} 

\maketitle
 \thispagestyle{fancy}
\pagestyle{fancy}
\lhead{This paper appears in IEEE Global Communications Conference (GLOBECOM) 2023.}
\cfoot{\thepage}
\renewcommand{\headrulewidth}{0.4pt}
\renewcommand{\footrulewidth}{0pt}

\begin{abstract}
Semantic communication is deemed as a revolution of Shannon's paradigm in the six-generation (6G) wireless networks. It aims at transmitting the extracted information rather than the original data, which receivers will try to recover.
Intuitively, the larger extracted information, the longer latency of semantic communication will be. Besides, larger extracted information will result in more accurate reconstructed information, thereby causing a higher utility of the semantic communication system. Shorter latency and higher utility are desirable objectives for the system, so there will be a trade-off between utility and latency.
This paper proposes a joint optimization algorithm for total latency and utility.
Moreover, security is essential for the semantic communication system. We incorporate the secrecy rate, a physical-layer security method, into the optimization problem. The secrecy rate is the communication rate at which no information is disclosed to an eavesdropper. 
Experimental results demonstrate that the proposed algorithm obtains the best joint optimization performance compared to the baselines.
\end{abstract}



\vspace{-10pt}
\section{Introduction \vspace{-5pt}}

After a decade of evolution, communication needs have evolved from telephone conversations to ultra-low-latency video calls, virtual/augmented reality games, etc.
Some emerging demands and services have urgent requirements for low latency and low energy consumption in communication networks. 
Also, considering that the existing communication technologies have approached the limit of Shannon's physical-layer capacity, it is worth contemplating the characteristics and features that could define the next-generation communication network \cite{9679803}.
The ultimate goal of traditional data-oriented communication is to convey semantic information. Hence, semantic communication \mbox{(SeComm)} has attracted attention from scholars recently \cite{9679803,qin2021semantic}.  Instead of transmitting the complete original information, SeComm extracts the critical information from the original data for transmission. 

\textbf{The utility and latency for semantic communication under physical layer security}.
To ensure a superior user experience, latency serves as a critical metric. Intuitively, the larger the extracted information of the original data, the more accurate the reconstructed data will be. 
However, larger extracted data means the time used for computation and transmission will be longer and causes long latency. 
In this paper, we define the utility of SeComm as how much data is recovered from the original data. Note that the larger utility, the longer latency. Nevertheless, for SeComm, it is indisputable that large utility and reduced latency are both critical. Therefore, we formulate a joint optimization problem, which minimizes the total latency (i.e., the sum of computation and transmission time) and maximizes the utility of SeComm.
Moreover, security is critical to SeComm since the extracted information should only be available to the intended receivers.
Thus, instead of the original transmission rate, we incorporate the secrecy rate into the formulated optimization problem.
The secrecy rate is a physical layer security method designed to prevent information leakage to eavesdroppers.

\textbf{Challenges}. First, SeComm is expected to replace traditional communication, thus sensitive to latency. Nevertheless, the communication resources (e.g., computing power, bandwidth) are often limited. 
Second, latency and utility are essential for the SeComm system. Hence, determining how to define latency and utility for the system and formulating the optimization problem is worth considering.
Third, eavesdroppers threaten physical layer security, so incorporating secrecy rate into the formulated problem is another challenge.

\textbf{Related Work}.
Recently, there have been several studies focusing on developing SeComm architectures. Several studies devised SeComm architectures for speech transmission \cite{weng2021sem_speech, tong2021fl_sem, han2023sem_preseved}.
Besides, some work proposed SeComm communication methods for text transmission \cite{xie2021semantic, xie2021semantic2, wang2022semantic, 9791409}.
Furthermore, there are also studies standing in the perspective of image transmission \cite{huang2021dl_img_sem,  yang2021_ofdm_img_trans,huang2023_sem_img}.
Resource allocation, which is the focus of this work, is also a significant genre of problems worth investigating in SeComm systems \cite{yang2023energy,yan2022allo_sem, yan2022qoe_allo_sem, su2023sem_allo}. \cite{yang2023energy} investigated an optimization strategy for the SeComm system under non-orthogonal multiple access (NOMA) to minimize total energy consumption with the constraints of transmission power, latency and computation. \cite{yan2022allo_sem} defined the semantic spectral efficiency and proposed a resource allocation algorithm to optimize the channel allocation and transmitted semantic symbols.
\cite{yan2022qoe_allo_sem} devised a quality-of-experience model and formulated a strategy to optimize the transmitted semantic symbols, channel allocation and power. \cite{su2023sem_allo} proposed a resource allocation strategy for the device-to-device vehicular SeComm network. Unlike the previous work, we propose a resource allocation scheme for the SeComm system in frequency-division multiple access (FDMA) to allocate appropriate bandwidth, transmission power and the size of transmitted information.

\textbf{Contributions.} The main contributions are as follows:
\begin{itemize}
    \item To our best knowledge, we are the first to implement resource allocation for SeComm while guaranteeing physical layer security, which could improve the performance and security of the whole system.
    \item An optimization algorithm is proposed to jointly optimize overall latency and utility within the system, taking into consideration the requirements of different scenarios.
    \item Detailed comparative experiments, time complexity, solution quality and convergence analysis are provided to show the robustness and effectiveness of our method.
\end{itemize}

\vspace{-5pt}
\section{System Model\vspace{-5pt}}

\begin{figure}
    \centering
    \includegraphics[width =0.78\linewidth]{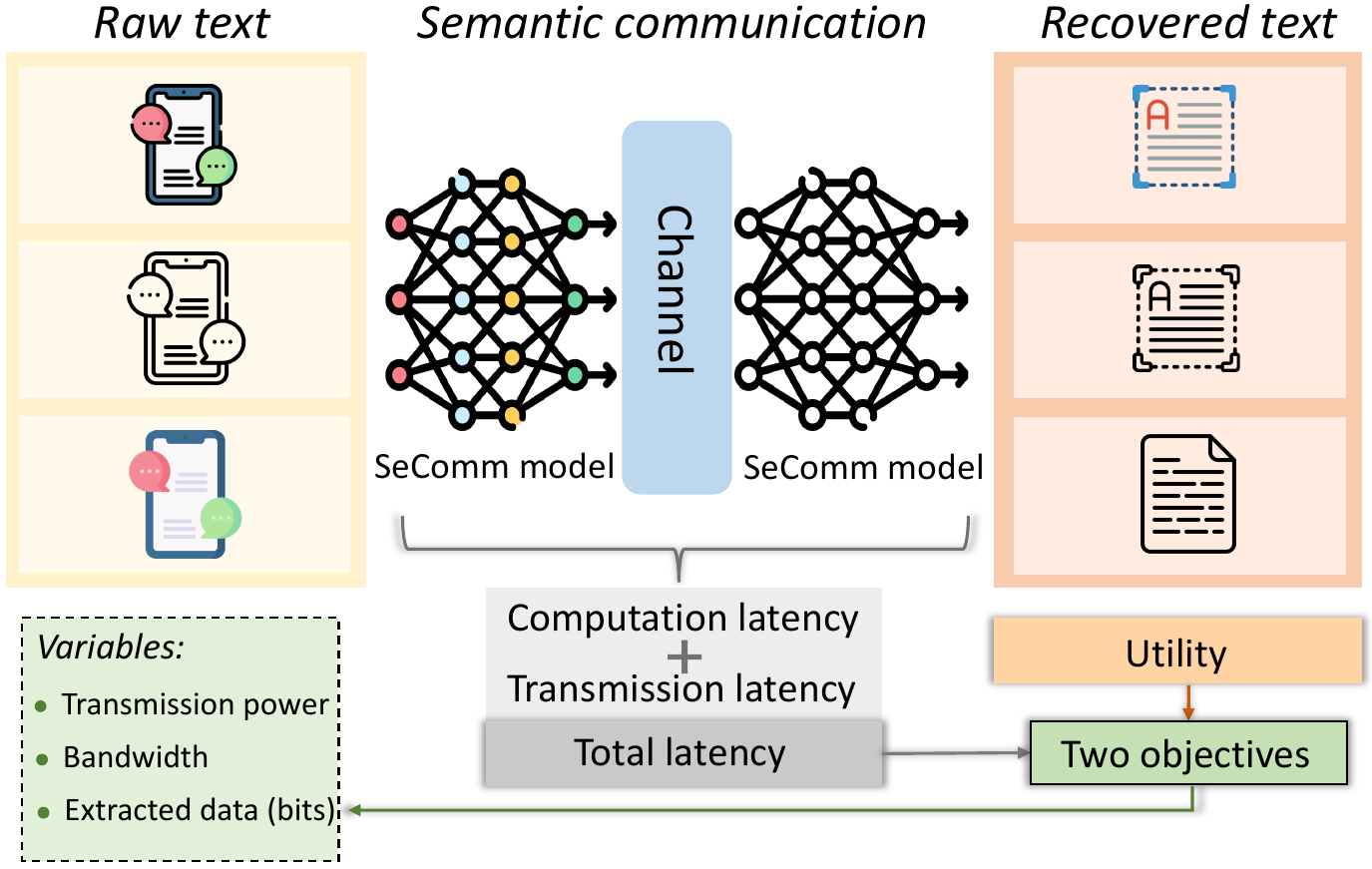}
    \caption{The semantic communication (SeComm) system model.}
    \label{fig:1}
\end{figure}

We consider an FDMA-based downlink SeComm system with $N$ users, as shown in Fig.\ref{fig:1}. Assume in the system that each user has $D_n$ bits of data to receive, and the server needs to extract semantic information of small-size $S_n$ (bits) from the original data for transmission due to limited wireless resources. 

\textbf{Semantic communication model.} 
A directional probability graph is utilized in our model to implement SeComm. The vertex in the graph denotes the semantic entity at different semantic levels, and the link between the two vertices represents the probability of association. Specifically, we can first use sequence labelling to identify the semantic entity and present it as a vertex. 
The subsequent phase involves utilizing a convolutional neural network, which is capable of capturing the interrelationships among entities, thereby enabling the calculation of the likelihood of a connection between any two vertices. 
Finally, semantic information fusion is conducted based on a constructed directional probability graph.

The extraction process aims to compress the size of transmission data in SeComm. A directional probability graph serves to extract semantic information with $S_n$ (bits) from the original data of size $D_n$. Then the semantic information is transmitted from the server to users via the downlink channel.

On the user side, each user $n$ will receive the transmitted semantic information with $S_n$ and utilizes the directional probability graph to retrieve the original data. 
Typically, a greater value of $S_n$ corresponds to a higher degree of preserved semantic information and facilitates a more precise retrieval of the original data.
The probability graph is shared among users to guarantee the consistency of recovered information.

\textbf{FDMA}. In this paper, we adopt FDMA technology. Assuming that the total available bandwidth is $B_{\textnormal{total}}$, each user $n$ would be allocated with $B_n$, and the constraint is $\sum_{n \in N} B_n \leq B_{\textnormal{total}}$. The server will communicate with users via different bandwidths, and we suppose there is no interference. In addition, the total assignable transmission power of the server is $p_{\textnormal{total}}$, so we have constraint $\sum_{n \in N} p_n \leq p_{\textnormal{total}}$, where $p_n$ is the downlink transmission power for user $n$.


\subsection{Total Time Consumption}
\textbf{Computation time of server:} 
The server is required to apply the SeComm model to process original data of size $D_n$ and extract semantic information of $S_n$ for each user $n$. The computation time $T_{1n}$ is defined as:
\begin{align}
    \textstyle{T_{1n} = \frac{y_{1n}(D_n,S_n)}{f_n},}
\end{align}
where function $y_{1n}(D_n, S_n)$ is to decide the required number of computation cycles for the retrieval of semantic information, and $f_n$ is the allocated computing capacity to user $n$ at the server. For  $y_{1n}(D_n,S_n)$, we utilize the following function proposed by\cite{yang2023energy} and propose a modified version:
\begin{align}
    &\textstyle{y_{1n}(D_n,S_n)} \!=\! \textstyle{y_{2n}(D_n)} \!+\! \textstyle{C_{1n}(\frac{S_n}{D_n}-1)^{C_{2n}}, }
\end{align}
where $C_{1n} > 0$, and $ C_{2n} $ is a positive even number. Both are constant parameters that could be obtained by function fitting in simulations. The first part $y_{2n}(D_n)$ is to compute the directional probability graph based on original data, which could be modelled as a concave function related to $D_n$. The second part refers to extracting semantic information of size $S_n$ (bits) from the graph. The smaller $\frac{S_n}{D_n}$ is, the more computing resources have to be used for better extraction. When $\frac{S_n}{D_n}=1$, the extraction scheme is straightforward, and the number of computation cycles is the lowest. Hence, the computation of the second part decreases as the ratio $\frac{S_n}{D_n}$ increases.


\textbf{Transmission time:}
According to the Shannon formula, the transmission rate $r_n(p_n,B_n)$ of legitimate user $n$ is:
\begin{align}
   \textstyle{ r_n(p_n,B_n) = B_n \log_2{\big(1+\frac{p_nh_n}{\sigma _n^2B_n}\big)}.}
\end{align}
Assume that for each user $n$, there exists a corresponding eavesdropper $E_n$ who aims to intercept the communication between the server and user $n$.
As FDMA is employed, the eavesdropping rate of $E_n$ is linked to the allocated bandwidth $B_n$, suggesting that eavesdropper $E_n$ could exploit $B_n$ to intercept user $n$'s information.
Let $r_{n,e}$ represents the eavesdropping rate of $E_n$ and we defined it as follows:
\begin{align}
    \textstyle{r_{n,e}(B_n) = B_n \log_2{\big(1+\frac{p_n^{eve}h_n^{eve}}{(\sigma_n^{eve})^2B_n}\big)},}
\end{align}
where $p_n^{eve}$, $h_n^{eve}$ and $(\sigma_n^{eve})^2$ are the transmission power, channel gain and noise density of the eavesdropper $E_n$ respectively. Thus, the secrecy rate of user $n$ is defined as:
\begin{align}
    \textstyle{r_{n,s}(p_n,B_n)} \!=\! \textstyle{r_n(p_n,B_n)}\! -\!\textstyle{ r_{n,e}(B_n).}
    \label{secrecy_rate}
\end{align}
Moreover, the transmission time $T_{2n}$ is defined as:
\begin{align}
   \textstyle{ T_{2n} = \frac{S_n}{r_{n,s}(p_n,B_n)},}
\end{align}
\vspace{-2pt}
\textbf{Computation time of user:}
After receiving the transmitted data, user $n$ needs to compute the semantic information with $S_n$ to recover the original data. So the computation time $T_{3n}$ could be defined as:
\begin{align}
    \textstyle{T_{3n} = \frac{y_{3n}(S_n)}{g_n},}
\end{align}
where function $y_{3n}(S_n)$ is to compute the number of CPU cycles needed to recover original data according to semantic information, and $g_n$ is the computation capacity of user $n$. 

We define $y_{3n}(S_n)$ like \cite{yang2023energy} as follows:
\begin{align}
    \textstyle{y_{3n}(S_n) = C_{3n}S_n^{-C_{4n}}},
\end{align}
where $C_{3n} > 0$ and $C_{4n} > 0$ are also constant parameters obtained in simulations.

\vspace{-5pt}\subsection{Utility of Semantic Information\vspace{-3pt}}
We also consider the utility of recovering the original information $D_n$ from the semantic information $S_n$ (i.e., the accuracy of semantic communication). 
It is intuitive that the larger the $S_n$, the easier it is to recover $D_n$ from $S_n$, so the utility function should be non-decreasing. Besides, there is a marginal effect on information recovery, so the function should be concave. The utility function is defined as follows:
\begin{align}
    U_n = y_{4n}(S_n)= 1-e^{-C_{5n} S_n},
\end{align}
where $C_{5n} \geq 0$ is also the constant parameter. Increased $S_n$ affects communication latency, whereas reduced $S_n$ compromises utility, so we need to ascertain an equilibrium.


\vspace{0pt}
\section{Joint optimization of Time and Utility\vspace{0pt}}
In this section, problem formulation, problem transformation and optimization algorithm will be introduced in detail. An analysis of the proposed method is also provided.
\vspace{-2pt}
\subsection{Problem Formulation\vspace{-2pt}}
A joint optimization problem considering both time consumption and utility is formulated as follows:
\begin{subequations}\label{Original_Problem}
\begin{align}
    \textnormal{Problem $\mathbb{P}_1$:~} \notag \\
    \min_{p_n,B_n,S_n} 
    \sum_{n \in N} &\big( \omega_1 (T_{1n}+T_{2n}+T_{3n}) - \omega_2 U_n \big),
    \tag{\ref{Original_Problem}} \\
    \text{subject to}:~
    &S_n \leq S_n^{max},~p_n^{min} \leq p_n,~n \in N, \label{constra:S_p} \\
    & \textstyle{\sum_{n \in N} p_n \leq p_{\textnormal{total}}},
    \label{constra:p_sum} \\
    & \textstyle{\sum_{n \in N} B_n \leq B_{\textnormal{total}}}, \label{constra:B_sum} 
\end{align}
\end{subequations}
where $p_n, B_n, S_n$ are optimization variables, $\omega_1$, and $\omega_2$ are weight parameters for latency and utility\footnote{Since latency and utility may have different scales, we utilize two weighting factors to control the optimization performance.}. Constraint (\ref{constra:S_p}) limits the upper size of $S_n$ to guarantee the effectiveness of the SeComm system and sets the range of transmission power for each user $n$.  $p_{\textnormal{total}}$ in constraint (\ref{constra:p_sum})  refers to the maximum available transmission power of the server. $B_{\textnormal{total}}$ in (\ref{constra:B_sum}) is the FDMA downlink bandwidth constraint. Moreover, condition \ref{condition:p} below is about minimum transmission power $p_n$.

\begin{condition}
\label{condition:p}
For all $n \in N$, $p_n^{min} \geq \frac{\sigma_n^2 h_n^{eve} p_n^{eve}}{(\sigma_n^{eve})^2 h_n}$
\end{condition}
\begin{remark}
Condition \ref{condition:p} with (\ref{constra:S_p}) ensures $r_n(p_n,B_n) \geq r_{n,e}(B_n)$, which means each user's secrecy transmission rate $r_{n,s}$ is non-negative according to (\ref{secrecy_rate}).
\end{remark}
\vspace{-8pt}
\begin{lemma}\label{lemma:rns}
$r_{n,s}(p_n,B_n)$ is neither convex nor concave.
\end{lemma}
\textit{Proof}.
Derive the Hessian matrix of $r_{n,s}$ and it is neither positive nor negative, so $r_{n,s}$ is neither convex nor concave.

\textbf{Difficulty of solving problem $\mathbb{P}_1$}. Note that $r_{n,s}$ in the objective function of $\mathbb{P}_1$ is neither convex nor concave, which makes the problem intractable. Also, the term $T_{2n}$, defined as $\frac{S_n}{r_{n,s}(p_n,B_n)}$, is a ratio. Thus, we need to minimize the sum of ratios in $\mathbb{P}_1$, and such a sum-of-ratio optimization problem is NP-complete~\cite{zhou2022joint}, which is challenging to solve.

\subsection{Successive Convex Approximation}
In this section, we employ the Successive Convex Approximation (SCA) technique \cite{razaviyayn2014successive} to handle neither convex nor concave function $r_{n,s}(p_n, B_n)$. Remember that 
$r_{n,s}(p_n,B_n)$ could be expressed as:
\begin{align}
     r_{n,s}(p_n,B_n) = r_n(p_n,B_n) - r_{n,e}(B_n),
    \label{constra:gamma1}
\end{align}
By using the first-order Taylor series to replace $r_{n,e}(B_n)$ in the second term, it could be approximated by:
\begin{align}
      \textstyle{r_{n,e}(B_n^{(i)})+\frac{\partial r_{n,e}}{\partial B_n}|_{B_n = B_n^{(i)}}(B_n-B_n^{(i)}),}
\end{align}
where the superscript $(i)$ means the value of the variable in $i$-th iteration. With the above approximation, we could rewrite $r_{n,s}(p_n, B_n)$ as follows:
\begin{align}
    \textstyle{r_n(p_n,B_n)- r_{n,e}(B_n^{(i)})-\frac{\partial r_{n,e}}{\partial B_n}|_{B_n = B_n^{(i)}}(B_n-B_n^{(i)})}.
\end{align}
Denoting $r_n(p_n,B_n)- r_{n,e}(B_n^{(i)})-\frac{\partial r_{n,e}}{\partial B_n}|_{B_n = B_n^{(i)}}(B_n-B_n^{(i)})$ by $R_n(p_n,B_n)$ for simplicity, then the following lemma holds:
\begin{lemma}
    $R_n(p_n,B_n)$ is a concave function.
\end{lemma}
\textit{Proof.}
    $r_n(p_n, B_n)$ is already a concave function which has been proved in Appendix A of \cite{zhou2022joint}. Remember that $B_n^{(i)}$ is the value of $B_n$ in $i$-th iteration, which can be regarded as a constant, so $- r_{n,e}(B_n^{(i)})-\frac{\partial r_{n,e}}{\partial B_n}|_{B_n = B_n^{(i)}}(B_n-B_n^{(i)})$ is actually an affine function. The summation of them is still a concave function based on Section 3.2 in \cite{boyd2004convex}.

Then we substitute $R_n(p_n,B_n)$ in $\mathbb{P}_1$ and rewrite it as:
\begin{subequations}\label{Problem_2}
\begin{align}
    \textnormal{Problem $\mathbb{P}_2$:~} \notag \\
    \min_{p_n,B_n,S_n} 
    \textstyle{\sum_{n \in N}} &
    \textstyle{\big(\omega_1 (T_{1n}+\frac{S_n}{R_n(p_n,B_n)}+T_{3n})\! - \! \omega_2 U_n\big)},
    \tag{\ref{Problem_2}} \\
    \text{subject to}:~& (\ref{constra:S_p}),~(\ref{constra:p_sum}),~
    (\ref{constra:B_sum}). \notag
\end{align}
\end{subequations}
Noting that $R_n(p_n,B_n)$ is concave now, the term $\frac{S_n}{R_n(p_n,B_n)}$ becomes a convex-concave ratio. Thus next, we manage to tackle the sum-of-ratio problem.
\vspace{-5pt}
\subsection{Transformation of the Sum-of-ratio Problem \vspace{-2pt}}

To tackle the sum-of-ratio optimization problem, we utilize the fractional programming technique proposed by \cite{zhao2023human} to transform problem $\mathbb{P}_2$ to equivalent problem $\mathbb{P}_3$:
\begin{subequations}\label{Problem_3}
\begin{align}
    &\hspace{-25pt} \textnormal{Problem $\mathbb{P}_3(\bm{z})$:~} \notag \\
    \min_{p_n,B_n,S_n} &
    \sum_{n \in N} 
    \textstyle{\big( W_n(S_n) \hspace{-2pt}+\hspace{-2pt} \omega_1 \hspace{-2pt}\cdot \hspace{-2pt}(S_n^2 z_n \hspace{-2pt} + \hspace{-2pt}\frac{1}{4 [R_n(p_n,B_n)]^2 z_n}) \big)},
    \tag{\ref{Problem_3}} \\
    &\hspace{-20pt}\text{subject to}:~
    (\ref{constra:S_p}),~
    (\ref{constra:p_sum}),~
    (\ref{constra:B_sum}), \notag
\end{align}
\end{subequations}
where we utilize $W_n(S_n)$ to denotes $\omega_1 (T_{1n}+T_{3n})-\omega_2U_n$, and 
introduce auxiliary variable $\bm{z}=[z_1,...,z_N]$ with $z_n \geq 0$. The process of using $\mathbb{P}_3$ to solve $\mathbb{P}_2$ is listed in Algorithm \ref{algo:fractional programming}.

\begin{algorithm}
\label{algo:fractional programming}
\caption{Fractional programming}
Initialize $j=0$, feasible $ \bm{X^{(0)}}= [ \bm{p^{(0)}}, \bm{B^{(0)}}, \bm{S^{(0)}}]$\\
Calculate 
$z_n^{(0)}$ = $\sqrt{\frac{G_n(\bm{X^{(0)}})}{F_n(\bm{X^{(0)}})}}$,
for $n=1,...,N.$\\
\Repeat{Convergence or reach max iteration number $J$}{

Obtain $\bm{X^{(j+1)}}$ by solving problem $\mathbb{P}_3$ according to Algorithm \ref{algo:KKT} in section \ref{sec:KKT} when given $\bm{z^{(j)}}$.

Update 
\begin{align}
    z_n^{(j+1)} = \textstyle{\sqrt{\frac{G_n(\bm{X^{(j+1)}})}{F_n(\bm{X^{(j+1)}})}},~\text{for } n=1,...,N. }\notag
\end{align}
Let $j \leftarrow j+1$.
}
\end{algorithm}

In Algorithm \ref{algo:fractional programming}, we denote the optimization variables $[\bm{p},\bm{B},\bm{S}]$ by $\bm{X}$, and write the objective function of Problem $\mathbb{P}_3$ as $\sum_{n\in N} \big(W_n(X) + F_n(X)z_n + \frac{G_n(X)}{z_n} \big)$, where $F_n(X)=S_n^2$ and $ G_n(X) = \frac{1}{4 [R_n(p_n,B_n)]^2}$ for simplicity. Through iterative solving and updating, we could obtain $\bm{z}$.

Until now, what we need to focus on is how to solve $\mathbb{P}_3$ when $\bm{z}$ is already derived from Algorithm \ref{algo:fractional programming}. Given conditions that functions $y_{1n}(D_n,S_n), y_{2n}(D_n,S_n)$ are convex, function $y_{3n}(S_n)$ is concave and constraints (\ref{constra:S_p}), (\ref{constra:p_sum}), (\ref{constra:B_sum}) are convex, $\mathbb{P}_3$ is a convex problem now. Thus, KKT conditions are sufficient and necessary to find the optimal solution.

\subsection{KKT conditions for Problem $\mathbb{P}_3$ }\label{sec:KKT}
We first write down the Lagrange function of $\mathbb{P}_3$:
\begin{align}
    &\textstyle{L = 
    \sum_{n \in N} \big(W_n(S_n)+ \omega_1 \cdot ({S_n}^2 z_n 
    + \frac{1}{4 [R_n(p_n,B_n)]^2 z_n})}\notag\\
    & \textstyle{ +\alpha_n\cdot( S_n-S_n^{max})  + \beta_n\cdot( p_n^{min}-p_n)
    \big) }\notag\\
    &\textstyle{
    +  \gamma \cdot (\sum_{n \in N} p_n - p_{\textnormal{total}}) 
    +\xi \cdot (\sum_{n \in N} B_n - B_{\textnormal{total}})},
\end{align}

After applying KKT conditions, we get:

\vspace{1pt}
\textbf{Stationarity: \vspace{1pt}}
\begin{subequations}
\begin{align}
    &\textstyle{\frac{\partial L}{\partial p_n}} \hspace{-2pt} =  \textstyle{-\frac{\omega_1}{4 R_n(p_n,B_n)^3 z_n}\frac{\partial R_n(p_n,B_n)}{\partial p_n}} \hspace{-2pt} -\hspace{-2pt} \beta_n \hspace{-2pt}+ \hspace{-2pt}\gamma = 0, \label{Lagrange:partial_p}\\
    &\textstyle{\frac{\partial L}{\partial B_n}}\hspace{-2pt} =  \textstyle{-\frac{\omega_1}{4 R_n(p_n,B_n)^3 z_n}\frac{\partial R_n(p_n,B_n)}{\partial B_n} + \xi = 0}, \label{Lagrange:partial_B}\\
    &\textstyle{\frac{\partial L}{\partial S_n}\hspace{-2pt} = \frac{\partial W_n(S_n)}{\partial S_n} + 2\omega_1S_nz_n  + \alpha_n = 0,} \label{Lagrange:partial_S}
\end{align}
\end{subequations}
\vspace{1pt}

\textbf{Complementary slackness:\vspace{1pt}}
\begin{subequations}
\begin{align}
    \text{(\ref{Complementary slackness}a):~}
    &\alpha_n \cdot (S_n-S_n^{max}) = 0,
    \text{(\ref{Complementary slackness}b):~}
    \beta_n \cdot (p_n^{min}-p_n) =0,\notag\\
    \text{(\ref{Complementary slackness}c):~}
    &\gamma \cdot (\sum_{n \in N} p_n\hspace{-2pt}-\hspace{-2pt} p_{\textnormal{total}})\hspace{-2pt} =\hspace{-2pt} 0, 
    \text{(\ref{Complementary slackness}d):~}
    \xi \cdot (\sum_{n \in N} B_n\hspace{-2pt} -\hspace{-2pt} B_{\textnormal{total}})\hspace{-2pt} =\hspace{-2pt} 0. \notag
\end{align}
\label{Complementary slackness}
\end{subequations}

\textbf{Primal feasibility: }(\ref{constra:S_p}),~(\ref{constra:p_sum}),~(\ref{constra:B_sum}).

\textbf{Dual feasibility:}
\begin{subequations}
    \text{(\ref{Dualfeasibility}a)-(\ref{Dualfeasibility}d):~}
    $\alpha_n,~\beta_n,~\gamma,~\xi \geq 0.$  
\label{Dualfeasibility}
\end{subequations}

With the above conditions, we try to identify a roadmap to derive the optimal variables step-by-step. Specifically, \textbf{Theorem \ref{theorem:KKT}} and \textbf{Algorithm \ref{algo:KKT}} are given to obtain the optimal solution.

\begin{theorem}\label{theorem:KKT}
    The optimal solution $\bm{p}^*,\bm{B}^*,\bm{S}^*$ could be derived from \textbf{Algorithm \ref{algo:KKT}} and expressed as:
    \begin{align}
    \begin{cases}
        {p}^*_n= \max\{\widehat{p}_n(\xi^*,\gamma^* \mid \boldsymbol{z}),p_n^{min}\} \\
        {B}_n^* = \widehat{B}_n(p_n^*, \xi^*\mid \boldsymbol{z}) \\
        S_n^* = \min \{ \widehat{S}_n(\bm{z}), S_n^{max} \}
    \end{cases}
    \end{align}
    where variable $\bm{z}$ is derived from Algorithm \ref{algo:fractional programming}, $\gamma^*$ and $\xi^*$ are obtained by using bisection method in Algorithm \ref{algo:KKT}.
\end{theorem}
\textit{Proof.}
    Please see Appendix \ref{Appendix:KKT}.

\begin{algorithm}
\label{algo:KKT}
\caption{Solve KKT Conditions}
Given auxiliary variable $\bm{z}$  \\

\For{$n \leftarrow 1$ to $N$}
{
Solve condition (\ref{Lagrange:partial_B}) and obtain $\widehat{B}_n(p_n, \xi \mid \boldsymbol{z})$.

Substituting $B_n = \widehat{B}_n(p_n, \xi \mid \boldsymbol{z})$, $ \beta_n = 0$ in (\ref{Lagrange:partial_p}), solve it and derive the solution $\widehat{p}_n(\xi,\gamma \mid \bm{z})$.

$\widetilde{p}_n(\xi,\gamma \mid \bm{z}) \leftarrow \max\{\widehat{p}_n(\xi,\gamma \mid \bm{z}), p_n^{min}\}$.

$\widetilde{B}_n(\xi,\gamma\mid \boldsymbol{z}) \leftarrow \widehat{B}_n(\widetilde{p}_n(\xi,\gamma\mid \boldsymbol{z}), \xi,\mid \boldsymbol{z})$.
}

$\gamma^*\hspace{-5pt} \leftarrow \hspace{-4pt}\begin{cases}
    0 ,~\text{if $\sum_{n \in N} \widetilde{p}_n(\xi,0 \hspace{-2pt} \mid \hspace{-2pt} \boldsymbol{z}) \leq p_{\textnormal{total}}$}\\\text{Solution to}
    \sum_{n \in N} \widetilde{p}_n(\xi,\gamma \hspace{-2pt} \mid \hspace{-2pt}  \boldsymbol{z})\hspace{-2pt}=\hspace{-2pt} p_{\textnormal{total}},\hspace{-2pt} \text{otherwise}
\end{cases}$

$\xi^*\hspace{-5pt}\leftarrow \hspace{-4pt}
\begin{cases}
    0 ,~\text{if $\sum_{n \in N} \widetilde{B}_n(0,\gamma^* \hspace{-2pt}\mid \hspace{-2pt}\boldsymbol{z}) \leq B_{\textnormal{total}}$}\\\text{Solution to}
    \sum_{n \in N}\hspace{-2pt} \widetilde{B}_n(\xi,\gamma^*\hspace{-3pt}\mid \hspace{-2pt}\boldsymbol{z})\hspace{-2pt} =\hspace{-2pt} B_{\textnormal{total}},\hspace{-2pt}\text{otherwise}
\end{cases} $


\textbf{Update }${p}^*_n \leftarrow \max\{\widehat{p}_n(\xi^*,\gamma^* \mid \bm{z}), p_n^{min}\}$, \\
~~~~~~~~~~${B}_n^* \leftarrow \widehat{B}_n(p_n^*,\xi^*\mid \boldsymbol{z})$.

\For{$n \leftarrow 1$ to $N$}{Setting $\alpha_n=0$ in condition (\ref{Lagrange:partial_S}), solve it and obtain the solution of $S_n$ as $\widehat{S}_n(\bm{z})$.

$S_n^* \leftarrow \min\{\widehat{S}_n(\bm{z}),S_n^{max}\}$.
}
\textbf{Return} $\bm{p}^*,\bm{B}^*,\bm{S}^*$ as the optimal solution.
\end{algorithm}

\vspace{-5pt}
\subsection{Resource Allocation Algorithm\vspace{-2pt}}

This section gives the complete resource allocation algorithm in \textbf{Algorithm \ref{algo:Resource_allocation}}.
It first initializes a feasible solution according to constraints (\ref{constra:S_p})-(\ref{constra:B_sum}),
 and implements the SCA method to transform Problem $\mathbb{P}_1$ into $\mathbb{P}_2$. Then it repeatedly utilizes Algorithm \ref{algo:fractional programming} and Algorithm \ref{algo:KKT} to update the values of $\bm{z}$ and $(\bm{p},\bm{B},\bm{S})$ until convergence. A globally optimal solution could be found for $\mathbb{P}_3$ which is equivalent to $\mathbb{P}_2$.

\begin{algorithm}
\label{algo:Resource_allocation}
\caption{Resource Allocation Algorithm}
Initialize feasible solution $sol^{(0)}=(\bm{p}^{(0)},\bm{B}^{(0)},\bm{S}^{(0)})$ of Problem $\mathbb{P}_1$, iteration number $k=1$\\

Implement SCA method to obtain $B_n^{(i)}$ and transform Problem $\mathbb{P}_1$ into $\mathbb{P}_2$ 

\Repeat{ $|sol^{(k)}-sol^{(k-1)}| \leq \epsilon_0 $ or reaching the maximum iteration number K}
{

Formulating $\mathbb{P}_3$, utilize Algorithm \ref{algo:fractional programming} and obtain $\bm{z}^{(k)}$ based on given $(\bm{p}^{(k-1)},\bm{B}^{(k-1)},\bm{S}^{(k-1)})$

Solve Problem $\mathbb{P}_3$ through Algorithm \ref{algo:KKT} and obtain solution 
$(\bm{p}^{(k)},\bm{B}^{(k)},\bm{S}^{(k)})$
based on $\bm{z}^{(k)}$\\
$sol^{(k)}=(\bm{p}^{(k)},\bm{B}^{(k)},\bm{S}^{(k)})$
\\
Set $k \leftarrow k+1$.
} 

\textbf{Return} $\bm{p}^{(k)},\bm{B}^{(k)},\bm{S}^{(k)}$ as the optimal solution
\end{algorithm}
\vspace{-5pt}

\subsection{Time complexity, Solution Quality and Convergence\vspace{-2pt}}

\textbf{Time Complexity. }
The complexity of Algorithm \ref{algo:Resource_allocation} lies in steps 2-8. The complexity of utilizing SCA in step 2 is $\mathcal{O}(IN)$, where $I$ is the number of iterations and $N$ derives from computing $B_n^{(i)}$ for each user in an iteration. 
In step 4, Algorithm \ref{algo:fractional programming} calls Algorithm \ref{algo:KKT}, so we first turn the view to the latter. 
Computing  $\widehat{p}_n$, $\widehat{B}_n$ and $\widehat{S}_n$ takes $\mathcal{O}(N)$ separately. Thus the total complexity of Algorithm \ref{algo:KKT} is $\mathcal{O}(3N)$. 
Assuming $J$ is the number of iterations in Algorithm \ref{algo:fractional programming}, and step 5 of it costs $\mathcal{O}(N)$, the total complexity of Algorithm \ref{algo:fractional programming} is $\mathcal{O}(4JN)$. Let $K$ denote the number of iterations in Algorithm \ref{algo:Resource_allocation}, and we could derive the overall complexity $\mathcal{O}((4JK+3K+I)N)$.

\textbf{Solution quality and convergence. }Algorithm \ref{algo:Resource_allocation} mainly comprises SCA , Algorithm \ref{algo:fractional programming} and Algorithm \ref{algo:KKT}. SCA transforms $\mathbb{P}_1$ to $\mathbb{P}_2$ could result in some loss of optimality. However, Algorithm \ref{algo:fractional programming} is based on Dinkelbach's transform and Algorithm \ref{algo:KKT} is solving the KKT conditions, both without loss of optimality. Thus, steps 3-8 of Algorithm \ref{algo:Resource_allocation} can guarantee finding the global optimal solution for $\mathbb{P}_2$. The convergence of Algorithm \ref{algo:Resource_allocation} is also evident from the preceding analysis.

\section{Experimental results\vspace{-2pt}}
In this section, we evaluate the effectiveness of our proposed method. First, we present the experimental parameter settings and discuss experimental results in other subsections.
\vspace{-5pt}
\subsection{Parameter Settings\vspace{-2pt}}\label{section:settings}
In the experiments, $N$ denoting the number of users is $30$. The path loss model is $128.1+37.6\log_{10}(\text{distance})$ with the standard deviation of shadow fading  $8$ dB, and the unit of distance is kilometer. The power spectral density of Gaussian noise is $-174$ dBm/Hz. Total bandwidth $B_{\textnormal{total}}$ is $10$ MHz. The maximum assignable transmission power~$p_{\textnormal{total}}$~is $40$ dBm. 

Furthermore, the assigned computing capacity $f_n$ and computation capacity of the user $g_n$ are set as $10$ GHz and $2$ GHz by default. The minimum transmission power $p_n^{min}$ is $0$ dBm. $S_n^{max}$, the maximum size of semantic information is $30$ MB.

\begin{figure}[!t]
\centering
\includegraphics[width=0.48\textwidth]{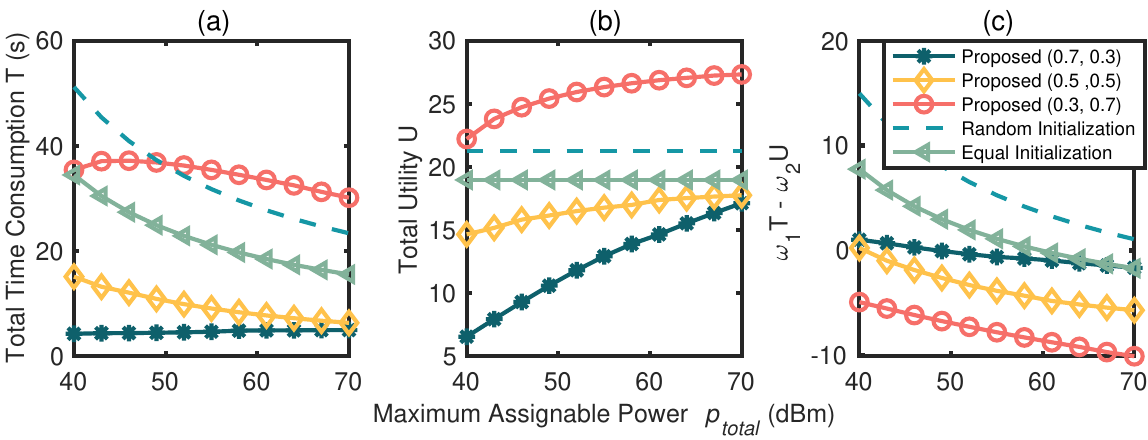}
\caption{Consumption under different maximum assignable power.}
\label{fig:Consumption under different maximum assignable power}
\end{figure}

\begin{figure}[!t]
\centering
\includegraphics[width=0.48\textwidth]{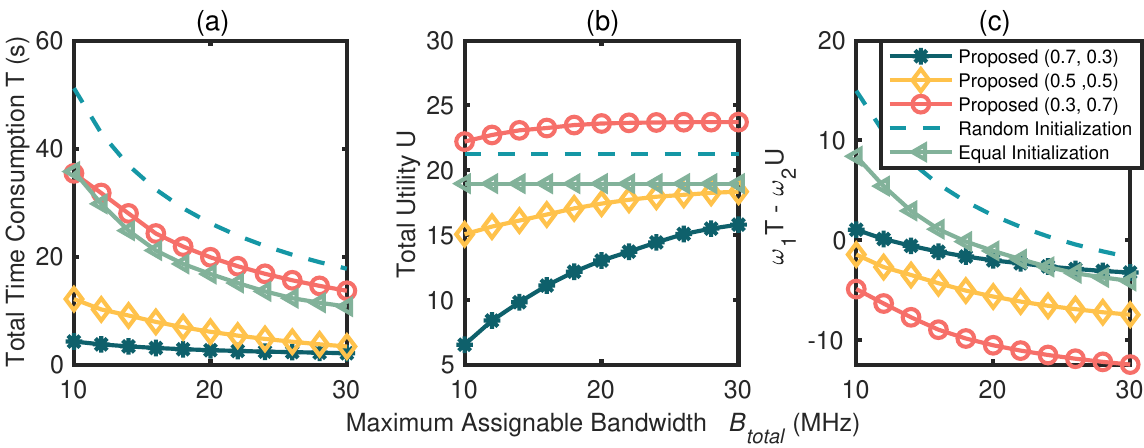}
\caption{Consumption under different maximum assignable bandwidth.  
}
\label{fig:Consumption under different maximum assignable bandwidth}
\end{figure}
\subsection{Performance when Adapting Weight Parameters\vspace{-2pt}}
We have two weighting factors $(\omega_1, \omega_2)$ to control the effect of optimization. For increased latency sensitivity, augment $\omega_1$. For superior utility, elevate $\omega_2$. To further investigate the influence of weight parameters, we conduct experiments under different combinations of $(\omega_1, \omega_2)$ and compare our method with two baselines: random and equal initialization. Random initialization means a random allocation of power and bandwidth after a random selection of $S_n$. Equal initialization distributes resources evenly to each user. We assume random and equal initialization have no preference, so their weight parameters are $(0.5,0.5)$.

Fig. \ref{fig:Consumption under different maximum assignable power}(a), (b), and (c) show the time consumption $T$, total utility $U$ and $\omega_1 T -\omega_2 U$ under diverse maximum assignable power $p_{\textnormal{total}}$. From Fig. \ref{fig:Consumption under different maximum assignable power}(a), we can see a decreasing trend in total time consumption as $p_{\textnormal{total}}$ increases, which is reasonable because more power could be allocated for transmission. 
The red line (Proposed (0.3,0.7)) performs worse than equal initialization on $T$, but the gap is smaller than that of $U$. It is also intuitive that as $p_{\textnormal{total}}$ grows, $U$ also increases. The reason is that increased assignable power means latency is no longer the bottleneck, and increasing $S_n$ leads to higher $U$, therefore, better joint optimization performance. In Fig. \ref{fig:Consumption under different maximum assignable power}(c), we can see that all proposed methods outperform the baselines, demonstrating our method's superiority.


Moreover, Fig. \ref{fig:Consumption under different maximum assignable bandwidth} demonstrates how our method performs as the assignable bandwidth $B_{\textnormal{total}}$ increases. The increasing trend $B_{\textnormal{total}}$ is similar to that of $p_{\textnormal{total}}$. Proposed methods contribute less latency than baselines in Fig. \ref{fig:Consumption under different maximum assignable bandwidth}(a). Although the $U$ of baselines are higher than our approach when not focusing on it (e.g., $\omega_1 = 0.3, \omega_1 = 0.5$), our method still generally outperforms the baselines, especially when  $B_{\textnormal{total}}$ is limited. 


\vspace{-5pt}
\subsection{Performance when Setting Different $S_n^{max}$\vspace{-3pt}}

To further investigate the performance of the proposed method, we consider diverse scenarios when the system requires different $S_n^{max}$. We fix the weight parameters $(\omega_1,\omega_2)=(0.3,0.7)$ to facilitate a better view of the relationship between algorithm performance and $S_n^{max}$. Four groups of experiments with different $p_{\textnormal{total}}$ and $B_{\textnormal{total}}$ are conducted.

\begin{figure}[!t]
\centering
\includegraphics[width=0.3\textwidth]{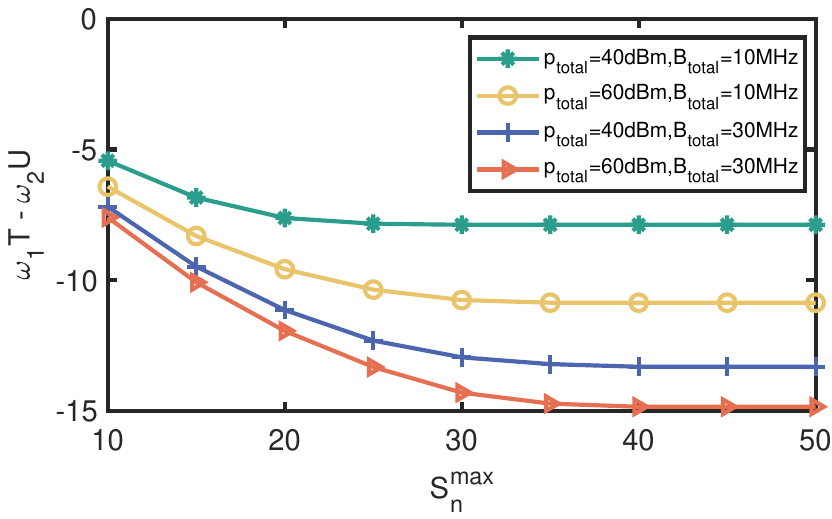}
\vspace{-10pt}
\caption{Joint optimization performance with different $S_n^{max}$.  
}
\label{fig:Smax}
\end{figure}

In Fig \ref{fig:Smax}, we can see that as $S_n^{max}$ increases, all four lines show a decreasing trend at the beginning and stabilize when finding the optimal $S_n$. It is also worth noting that 
when $p_{\textnormal{total}}$ and $B_{\textnormal{total}}$ are set higher, the algorithm demands a larger $S_n^{max}$ to converge. 
This is because the more communication resource a system has, the more it tends to choose higher $S_n$ to increase $U$ for better joint optimization performance.\vspace{-5pt}



\section*{Acknowledgement\vspace{-5pt}}

This research is partly supported by the Singapore Ministry of Education Academic Research Fund under Grant Tier 1 RG90/22, Grant Tier 1 RG97/20, Grant Tier 1 RG24/20 and Grant Tier 2 MOE2019-T2-1-176; and partly by the Nanyang Technological University (NTU)-Wallenberg AI, Autonomous Systems and Software Program (WASP) Joint Project.\vspace{-5pt}

\section{Conclusion\vspace{-5pt}}
In this paper, a downlink SeComm system that considers physical-layer security is proposed, to ensure secure and efficient transmission. By introducing different weight parameters, we also propose an adaptive resource allocation algorithm to jointly optimize the time consumption and utility of the whole system. A globally optimal solution could be derived, and experiments show the effectiveness of our method. 



\begin{appendices}



\section{Analysis of KKT Conditions}
\label{Appendix:KKT}
To find optimal $(\bm{p}^*,\bm{B}^*,\bm{S}^*)$ satisfying KKT conditions, the following analysis based on optimization theory is provided. 

\textbf{Analysis of \bm{$(\bm{p}, \bm{B})$}: }
Remember that $\bm{z}$ is already given, so we could derive the solution of $B_n$ represented by $(p_n,\xi)$ from (\ref{Lagrange:partial_B}) defined as $\widehat{B}_n(p_n, \xi \mid \boldsymbol{z})$.
Moreover, we can also obtain the following equation from (\ref{Lagrange:partial_p}) by setting $\beta_n = 0$:
\begin{align}
    \frac{-\omega_1}{4 R_n(p_n, B_n)^3 z_n}\frac{\partial R_n(p_n,B_n)}{\partial p_n}\hspace{-2pt}  + \hspace{-2pt} \gamma|_{B_n=\widehat{B}_n(p_n, \xi\mid \boldsymbol{z})}\hspace{-2pt} =\hspace{-2pt} 0, \label{p_n,gamma}    
\end{align}
The solution of $p_n$ in (\ref{p_n,gamma}) could be denoted as $\widehat{p}_n(\xi,\gamma\mid \boldsymbol{z})$, and
a discussion of it  with different conditions is given below:
\begin{itemize}
    \item \textbf{Case 1:} $\widehat{p}_n(\xi,\gamma\mid \boldsymbol{z}) \geq p_n^{min}$.
    In this case, we simply set $\beta_n = 0$ and $p_n = \widehat{p}_n(\xi,\gamma\mid \boldsymbol{z})$. 
    \item \textbf{Case 2:} $\widehat{p}_n(\xi,\gamma\mid \boldsymbol{z}) < p_n^{min}$.    
    It is obvious that we cannot set $\beta_n =0$ in this case, otherwise the solution of (\ref{Lagrange:partial_p}) will be $\widehat{p}_n(\xi,\gamma\mid \boldsymbol{z})$, which will violate  condition (\ref{constra:S_p}). Since $\beta_n >0$, $p_n = p_n^{min}$  could be derived from (\ref{Complementary slackness}b). Substituting $p_n = p_n^{min}$ in (\ref{Lagrange:partial_p}), we have $\beta_n = (-\frac{1}{4R_n(p_n,B_n)^3 z_n} \frac{\partial R_n(p_n,B_n)}{\partial p_n} + \gamma)\mid_{p_n=p_n^{min},B_n=\widehat{B}_n(p_n, \xi \mid \boldsymbol{z})} > 0$ due to the convexity of the problem. Thus $p_n = p_n^{min}$ could be set as the solution.
\end{itemize} 

Summarize both cases and the conclusion could be derived:
\begin{align}\label{p,B}
\begin{cases}
    \widetilde{p}_n(\xi,\gamma\mid \boldsymbol{z})= \max\{\widehat{p}_n(\xi,\gamma\mid \boldsymbol{z}),p_n^{min}\} \\
    \widetilde{B}_n(\xi,\gamma\mid \boldsymbol{z}) = \widehat{B}_n(\widetilde{p}_n(\xi,\gamma\mid \boldsymbol{z}), \xi,\mid \boldsymbol{z})
\end{cases}
\end{align}


Compute the value of $\sum_{n \in N} \widetilde{p}_n(\xi,0\mid \boldsymbol{z})$ by setting $\gamma=0$, and then discuss the following two cases:
\begin{itemize}
    \item \textbf{Case 1:} $\sum_{n \in N} \widetilde{p}_n(\xi,0\mid \boldsymbol{z}) \leq p_{\textnormal{total}}$.
    Set $\gamma = 0$ and conditions (\ref{constra:p_sum}), (\ref{Complementary slackness}c),  (\ref{Dualfeasibility}c) are satisfied.
    \item \textbf{Case 2:} $\sum_{n \in N} \widetilde{p}_n(\xi,0\mid \boldsymbol{z}) > p_{\textnormal{total}}.$ 
    Set $\gamma > 0$, and we can obtain from (\ref{Complementary slackness}c) that
    $\sum_{n \in N}\widetilde{p}_n(\xi,\gamma\mid \boldsymbol{z}) = p_{\textnormal{total}}$.
    A bisection method could be utilized to solve it and derive the solution as $\widehat{\gamma}(\xi \mid z)$.
\end{itemize}
\begin{align}\label{optimal:gamma}
\widetilde{\gamma}(\xi \mid \boldsymbol{z}) = 
\begin{cases}
    0 , &\text{if $\sum_{n \in N} \widetilde{p}_n(\xi,0\mid \boldsymbol{z}) \leq p_{\textnormal{total}}$}\\
    \widehat{\gamma}(\xi \mid \boldsymbol{z}) , &\text{otherwise}
\end{cases}
\end{align}

Similarly, compute the value of $\sum_{n \in N} \widetilde{B}_n(0,\widetilde{\gamma}(0)\mid \boldsymbol{z})$ by setting $\xi=0$ and then discuss two cases:
\begin{itemize}
    \item \textbf{Case 1:} $\sum_{n \in N} \widetilde{B}_n(0,\widetilde{\gamma}(0 )\mid \boldsymbol{z}) \leq B_{\textnormal{total}}$

    In this case, we set $\xi = 0$ and remaining KKT conditions (\ref{constra:B_sum}), (\ref{Complementary slackness}d),  (\ref{Dualfeasibility}d) are all satisfied.

    \item \textbf{Case 2:} $\sum_{n \in N} \widetilde{B}_n(0,\widetilde{\gamma}(0 )\mid \boldsymbol{z}) > B_{\textnormal{total}}$

    If we set $\xi =0$, condition (\ref{constra:B_sum}) will be violated. So we set $\xi > 0$ and obtain from (\ref{Complementary slackness}d) that
    $\sum_{n \in N} \widetilde{B}_n(\xi,\widetilde{\gamma}(\xi)\mid \boldsymbol{z}) = B_{\textnormal{total}}$. The solution of it is $\widehat{\xi}(z)$.
\end{itemize}
\begin{align}
\xi^*(\boldsymbol{z}) =
\begin{cases}
    0 , &\text{if $\sum_{n \in N} \widetilde{B}_n(0,\widetilde{\gamma}(0 )\mid \boldsymbol{z}) \leq B_{\textnormal{total}}$}\\
    \widehat{\xi}(\boldsymbol{z}) , &\text{otherwise}
\end{cases}
\end{align}

Substituting $\xi^*(z)$ to (\ref{p,B}) and (\ref{optimal:gamma}), we could obtain:
\begin{align}\label{optimal:p,B}
\begin{cases}
    {p}^*_n= \max\{\widehat{p}_n(\xi^*(\boldsymbol{z}),\widetilde{\gamma}(\xi^*(\boldsymbol{z}))\mid \boldsymbol{z}),p_n^{min}\} \\
    {B}_n^* = \widehat{B}_n(p_n^*, \xi^*(\boldsymbol{z})\mid \boldsymbol{z})
\end{cases}
\end{align}

\textbf{Analysis of $\bm{S}$: }
For the variable $S_n$, we first write down the following equation from (\ref{Lagrange:partial_S}) by setting $\alpha_n = 0$:
\begin{align}\label{s_n}
    \textstyle{\frac{\partial W_n(S_n)}{\partial S_n} + 2\omega_1S_nz_n  = 0.}
\end{align}

The solution $S_n$ of (\ref{s_n}) could be derived as $\widehat{S}_n(\bm{z})$. Then we do a comparison between $\widehat{S}_n(\bm{z})$ and $S_n^{max}$ and discuss:

\begin{itemize}
    \item \textbf{Case 1:} $\widehat{S}_n(\bm{z}) \leq S_n^{max}$

    We could simply set $\alpha_n = 0$ in this case, and $S_n = \widehat{S}_n(\bm{z})$ would be the solution of (\ref{Lagrange:partial_S}).
    \item \textbf{Case 2:}
    $\widehat{S}_n(\bm{z}) > S_n^{max}$

    Setting $\alpha_n > 0$,  we have $S_n = S_n^{max}$ from (\ref{Complementary slackness}a) and $\alpha_n = -( \frac{\partial W_n(S_n)}{\partial S_n} + 2S_nz_n) \mid_{S_n=S_n^{max}}>0$. 
\end{itemize}

\begin{align}
S_n^* = \min \{\widehat{S}_n(\bm{z}), S_n^{max} \}
\end{align}

Until now, all optimal variables $\bm{[p^*,B^*,S^*]}$ are derived.

\end{appendices}

\end{document}